\documentclass[review,12pt]{elsarticle}
\usepackage{lineno,hyperref}
\modulolinenumbers[5]
\usepackage{url}
\journal{Journal of \LaTeX\ Templates}

\usepackage{amsmath}
\usepackage{graphicx}
\usepackage{mathtools}

\title{ Hot electrons injection in carbon nanotubes under the influence of 
quasi-static ac-field   }

\author[els]{M. Amekpewu\corref{cor1}}
\author[rvt]{S. Y. Mensah}
\author[els]{R. Musah}
\author[focal]{ N. G. Mensah}
\author[rvt]{\\ S. S. Abukari}
\author[rvt]{ K. A. Dompreh}
\address[els]{Department of Applied Physics, University for Development Studies,
Navrongo, Ghana}
\address[rvt]{Department of Physics, College of Agriculture and Natural Sciences, U.C.C, Ghana.}
\address[focal]{Department of Mathematics, College of Agriculture and Natural Sciences, U.C.C, Ghana}
\cortext[cor1]{Corresponding author\\ Email: mamek219@gmail.com\\
This work is licensed under the Creative Commons Attribution International\\ License (CC BY).
\url{http://creativecommons.org/licenses/by/4.0/} }
\ead[url]{mamek219@gmail.com}

\date{}

\begin{document}
\begin{abstract}
Hot electrons injection in carbon nanotubes  (CNTs ) where in addition to  
applied dc field ($\mathbf{E}$), there exist simultaneously a quasi-static 
ac electric field (i.e. when the frequency $\omega$ of ac field is much less than 
the scattering frequency $v$ ($\omega\ll v$ or $\omega\tau\ll 1$, $v =\tau^{-1}$,
where $\tau$ is relaxation time) is considered. The investigation is done 
theoritically by solving semiclassical Boltzmann transport equation with and 
without the presence of the hot electrons source to derive the current densities. 
Plots of the normalized current density versus dc field ($\mathbf{E}$) applied 
along the axis of the CNTs in the presence and absence of hot electrons  reveal 
ohmic conductivity initially and finally negative differential conductivity (NDC) 
provided $\omega\tau\ll 1$ (i.e. quasi- static case).  With strong enough axial 
injection of the hot electrons , there is a switch from NDC to positive 
differential conductivity (PDC)  about  $\mathbf{E} \geq 75 kV/cm$ and 
$\mathbf{E} \geq 140 kV/cm$ for a zigzag CNT and an armchair CNT respectively. 
Thus, the most important tough problem for NDC region which is the space charge 
instabilities can be suppressed due to the switch from the NDC behaviour to the 
PDC behaviour predicting a potential generation of terahertz radiations whose 
applications are  relevance in  current-day technology, industry, and research.

\end{abstract}

\maketitle
\section*{Introduction}
Carbon nanotubes (CNTs)~\cite{1} -~\cite{3} are subject of many theoretical
~\cite{4} -~\cite{9}, and experimental~\cite{10} -~\cite{18} studies. 
Their properties include a thermal conductivity higher than diamond, 
greater mechanical strength than steel and better electrical conductivity than 
copper~\cite{19} -~\cite{21}. These novel properties  make them potentially useful 
in a variety of applications in nanotechnology, optics, electronics, and other 
fields of materials science~\cite{22}~\cite{24}. Rapid development of submicrometer 
semiconductor devices, which may be employed in high-speed computers and 
telecommunication systems,  enhances the importance of hot-electron phenomena~\cite{25}. 
Hot electron  phenomena  have  become  important  for the  understanding  of  all  
modern  semiconductor  devices~\cite{26}-~\cite{27}. There are several reports on hot 
electrons generation in CNTs~\cite{28}-~\cite{30}, but the reports on hot electrons 
injection in CNTs under the influence of quasi-static ac field to the best of our 
knowlege are limited. Thus, in this paper, we analyzed theoretically hot electrons 
injection in $(3,0)$ zigzag$(zz)$ CNT and $(3,3)$ armchair (ac) CNT  where in addition 
to dc field, a quasi-static ac electric field is applied. Adopting semiclassical 
approach , we obtained current density for each achiral CNTs after solving the 
Boltzmann transport equation  in the framework of momentum-independent relaxation 
time . We probe the behaviour of the electric current density of the CNTs as a 
function of the applied $dc$ field  $\mathbf{E}$ of $ac-dc$ driven fields when 
the frequency of ac field $(\omega)$  is much less than the scattering frequency ($v$) 
($\omega\ll v$ or $\omega\tau\ll1$ i.e quasi-static case~\cite{31}, where 
$v =\tau^{-1}$) with and without the axial injection of the hot electrons.

\section*{Theory}
Suppose an undoped single walled achiral carbon nanotubes (CNTs) $(n,0)$ or 
$(n,n)$ of length $L$ is exposed to a homogeneous axial dc field $\mathbf{E}$ 
given by  $\mathbf{E} = V/L$, where $V$ is the voltage between the CNT ends. 
Under the influence of the applied dc field and assuming scattering is negligible, 
electrons with electronic charge ($e$) obey Newton's law of motion given 
by~\cite{32}
\begin{equation}
\frac{d\mathbf{P}}{dt} = e\mathbf{E}
\end{equation}
where $\mathbf{P}$ is  a component of  quasimomentum along the  axis of the tube. 
Adopting semiclassical approximation approach and considering the motion of $\pi-$ 
electrons as a classical motion of free quasi-particles with dispersion law extracted 
from the quantum theory while taking  into  account to the hexagonal crystalline 
structure of CNTs and  applying the tight-binding approximation  gives the energies 
for $zz-$CNT and ac-CNT respectively

\begin{eqnarray}
\varepsilon(s\Delta{p_{\vartheta}},p) & \equiv &  \varepsilon_s(p)=\nonumber\\
&\pm &\gamma_0\left[{1 + 4cos\left(a{p}\right)cos\left(\frac{a}{\sqrt{3}}s\Delta p_{\vartheta}\right) + 4cos^2\left(\frac{a}{\sqrt{3}}s\Delta p_{\vartheta}\right)}\right]^{1/2}
\end{eqnarray}

\begin{eqnarray}
\varepsilon(s\Delta{p_{\vartheta}},p) &\equiv &  \varepsilon_s(p_z)=\nonumber\\
&\pm &\gamma_0\left[{1 + 4cos\left(as\Delta p\}\right)cos\left(\frac{a}{\sqrt{3}} p_{\vartheta}\right) + 4cos^2\left(\frac{a}{\sqrt{3}} p_{\vartheta}\right)}\right]^{1/2}
\end{eqnarray}
where $\gamma_{0}\approx3.0eV$ is the overlapping integral,  $\Delta p_{\vartheta}$ 
is transverse quasimomentum level spacing and $s$ is an integer.  The lattice constant 
$a$ in Eqn.($2$) and ($3$) is expressed as~\cite{33}
\begin{eqnarray}
a=\frac{3b}{2\hbar}
\end{eqnarray}                              
where $b=0.142nm$ is the C-C bond length . The $(-)$ and $(+)$ signs correspond to the 
valence and conduction bands respectively. Because of the transverse quantization of 
the quasimomentum $P$, its transverse component $p_{\vartheta}$ can take $n$ discrete 
values,
\begin{eqnarray}
p_{\vartheta}=s\Delta p_{\vartheta}=\frac{\pi\sqrt{3}s}{an}(s=1,.....,n)
\end{eqnarray}
As different from $p_{\vartheta}$, we assume $\mathbf{p}$  continuously varying within 
the range $0\le\mathbf{p}\le2\pi/a$  which corresponds to the model of infinitely long 
CNT $(L =\infty)$. The model is applicable to the case under consideration because we 
are restricted to temperatures and/or voltages well above the level spacing~\cite{33}, 
i.e. $k_{B} T>\varepsilon_{c}$, $\Delta\varepsilon$,  where $k_{B}$ is Boltzmann constant, 
$T$ is the thermodyanamic temperature, $\varepsilon_{c}$ is the charging energy. In the 
presence of hot electrons source, the motion of quasi-particles in an external axial 
electric field is described by the Boltzmann kinetic equation as~\cite{32}-~\cite{33}
\begin{equation}
\frac{\partial f(p)}{\partial t} + v\frac{\partial f(p)}{\partial x}+eE(t)\frac{\partial f(p)}{\partial p}  = -\frac{f(p)-f_{eq}(p)}{\tau} + S(p)
\end{equation}
where  $f_{eq}(p)$ is equilibrium Fermi distribution function, $f(p,t)$ is the 
distribution function, $S(p)$ is the hot electron source function, $\mathbf{v}$ 
is the quasiparticle group velocity along the axis of carbon nanotube  and $\tau$ 
is the relaxation time. The relaxation term of Eqn.($6$)  above describes the 
electron-phonon scattering, electron-electron collisions~\cite{34}~\cite{35} etc.\\

Applying the method originally developed in the theory of  quantum semiconductor 
superlattices~\cite{33}, an exact solution of equation (6) can be constructed 
without assuming a weak electric field. By expanding the distribution functions 
of interest in Fourier series, we have:
\begin{equation}
f(p,t) = \Delta p_{\vartheta} \sum_{s=1}^n{\delta(p_{\vartheta} - s\Delta p_{\vartheta})}
\sum_{r \neq 0}{f_{rs}exp({iarp})\psi_v(t)}
\end{equation}
and 
\begin{equation}
f_{eq}(p) = \Delta p_{\vartheta} \sum_{s=1}^n{\delta(p_{\vartheta} - s\Delta p_{\vartheta})}
\sum_{r \neq 0}{f_{rs}exp({iarp})}
\end{equation}
for $zz$-CNT and 
\begin{equation}
f(p,t) = \Delta p_{\vartheta} \sum_{s=1}^n{\delta(p_{\vartheta} - s\Delta p_{\vartheta})}
\sum_{r \neq 0}{f_{rs}\exp{ir(a/\sqrt{3}p)}\psi_v(t)}
\end{equation}
and\\
\begin{equation}
f_{eq}(p) = \Delta p_{\vartheta} \sum_{s=1}^n{\delta(p_{\vartheta} - s\Delta p_{\vartheta})}
\sum_{r \neq 0}f_{rs}\exp\{ir(a/\sqrt{3}p)\}
\end{equation}
for ac-CNTs\\
where $\delta(p_{\vartheta}-s\Delta p_{\vartheta})$ is the Dirac delta function, 
$f_{rs}$ is the coefficients of the Fourier series and $\psi_{v}(t)$ is the factor 
by which the Fourier transform of the nonequilibrium distribution function differs 
from its equilibrium distribution counterpart. For simplicity, we consider a hot 
electron source of the simplest form given by the expression,
\begin{equation}
S(p) = \frac{Qa}{\hbar}\delta(\varphi-\varphi^\prime) - \frac{aQ}{n_0}f_{s}(\varphi)
\end{equation}
where $f_s(p)$ is the static and homogeneous ( stationary) solution of Eqn.($6$),
$Q$ is the injection rate of hot electron, $n_0$ is the equilibrium particle density, 
$\varphi$ and $\varphi^\prime$ are the dimensionless momenta of  electrons and hot 
electrons respectively which are expressed as $\varphi = a\mathbf{p}/\hbar$ and 
$\varphi'= a\mathbf{p'}/\hbar$ for zz-CNTs and $\varphi = a\mathbf{p}/\sqrt{3}\hbar$ 
and $\varphi'= a\mathbf{p'}/\sqrt{3}\hbar$ for ac-CNTs,\\
We now obtain the current density in the nonequilibrium state for zz-CNT where 
in addition to  applied dc field, there exist simultaneously a quasi-static ac 
electric field by considering perturbations with frequency $\omega$ and wave-vector 
$\kappa$ of the form~\cite{32}. 
\begin{equation}
E(t) = \mathbf{E} + E_{\omega,k}exp(-i\omega t+ ikx)
\end{equation}
\begin{equation}
f =f_s(\varphi) + f_{\omega,k}exp(-i\omega t+ikx)
\end{equation}
where $\mathbf{E}$ is dc field along the axis of the tube,  $E_{\omega,\kappa} 
e^{-i\omega t+i\kappa x}$ is ac-field,  $E_{\omega,\kappa}$  is peak ac field and 
$f_{s}(\varphi)$is the static and homogeneous (stationary) solution of Eqn.($6$).
Substituting Eqn.($12$) and ($13$) into Eqn.($6$) and rearranging yields,           
\begin{equation}
\frac{\partial f_{\omega,k}}{\partial \varphi} + i[\alpha + {k\mathbf{v}_{z}\hbar}/{ae\mathbf{E}}]f_{\omega,k} = -\frac{E_{\omega,\kappa}}{\mathbf{E}}
\frac{\partial f_s(\varphi)}{\partial \varphi}
\end{equation}
where $\alpha = -\hbar(\omega + iv)/aeE$ and $f_{\omega,k}$ is the solution of 
Eqn.($14$). Solving the homogeneous differential Eqn.($14$)and and then 
introducing the Jacobi-Anger expansion, we obtain the normalized current density 
in the presence of hot electrons ($j_{HE}^{zz}$) as 
\begin{eqnarray}
j_{HE}^{zz} = i\frac{4\sqrt{3}e^2 \gamma_0}{n\hbar^2}
\sum_{r=1}{r} \sum_{s=1}{f_{rs}\varepsilon_{rs}}
\sum_{m,l=-\infty}\frac{i^l mlj_m(\beta)j_{m-l}(\beta)I_{m-l}(\beta)aeE}{([\omega + iv]\hbar - m(ae)\mathbf{E})}\nonumber\\
\times\left\{\eta\frac{n_{o}}{2\pi}\sum_{r}\frac{ae\mathbf{E}exp(ir\varphi)}{(ir(ae\mathbf{E}) + v\hbar +\eta(ae\mathbf{E}))}
\left.(exp(-ir\varphi^{\prime}) - \frac{v\hbar}{(v\hbar+ir(ae\mathbf{E}))}\right)\right.\nonumber\\
+\left.\frac{v\hbar}{(v\hbar+ir(ae\mathbf{E}))}\right\}
\end{eqnarray}
where\\
\begin{eqnarray*}
f_{rs}=\frac{a}{2\pi\Delta p_{\vartheta}}\int_{0}^{2\pi/a}
\frac{\exp(-iarp)}{1+\exp\{\varepsilon_{s}(p)/k_{B}T\}}dp
\end{eqnarray*}
\begin{eqnarray*}
\varepsilon_{rs}=\frac{a}{2\pi\gamma_{0}}\int_{0}^{2\pi/a}
\varepsilon_{s}(\mathbf{p})\exp(-iarp)dp
\end{eqnarray*}
$\beta=\kappa\gamma_{0}a/\Omega\hbar$, $\eta=Q/\Omega n_{0}$ and 
$\Omega=ea\mathbf{E}/\hbar$, $j_{m}(\beta)$ is the $m$th order Bessel function of 
the first kind, $J_{(m-1)}(\beta)$  is the $(m-1)$th order Bessel function of the 
first kind, $I_{(m-l)}(\beta)$is $(m-1)th$  order modified Bessel function of the 
first kind, $Q$  is rate of hot electrons injection, $n_{0}$  is the particle density 
$\Omega$ is the Bloch frequency and $\eta$ is the non-equilibrium parameter.\\

In the absence of hot electrons, the nonequalibrium parameter for $zz$-CNT,  
$\eta=0$, hence the current density for $zz$-CNTs without hot electron source 
$j^{zz}$ could be obtained from Eqn.($15$) by setting  $\eta=0$.  Therefore, 
the electric current density of $zz$-CNTs in the absence of hot  $j^{zz}$   is given by
\begin{multline}
j_{z}^{zz} = i\frac{4\sqrt{3}e^2 \gamma_0}{n\hbar^2} \sum_{l=1}{r} \sum_{s=1}{f_{rs}\varepsilon_{rs}}\\
\times\sum_{m,l=-\infty}{\frac{i^l mlj_m(\beta)j_{m-l}(\beta)I_{m-l}(\beta)(ae)\mathbf{E}}
{([\omega + iv]\hbar - m(ae)\mathbf{E})}}\left\{\sum_{r}\frac{v\hbar}{v\hbar+ir(ae\mathbf{E})}\right\}
\end{multline}
Applying similar argument like one for $zz$-CNT, the current density for an $ac$-CNT 
with and without the injection of hot electrons are expressed respectively as:
\begin{eqnarray}
j_{HE}^{ac} = i\frac{4e^2 \gamma_0}{\sqrt{3}n\hbar^2} \sum_{r=1} \sum_{s=1}{f_{rs}\varepsilon_{rs}}\sum_{m,l=-\infty}
{\frac{i^l mlj_m(\beta)j_{m-l}(\beta)I_{m-l}(\beta)(ae)\mathbf{E}}{(\sqrt{3}[\omega + iv]\hbar - m(ae)\mathbf{E})}}\nonumber\\
\times\left\{\eta\frac{n_{o}}{2\pi}\sum_{r}\frac{(ae)\mathbf{E} exp(ir\varphi)}{(ir(ae\mathbf{E})+\sqrt{3}v\hbar+\eta(ae\mathbf{E}))}
\left(exp({-ir\vartheta^{\prime}})-\frac{\sqrt{3}v\hbar}{(\sqrt{3}v\hbar+ir(ae\mathbf{E}))}\right)\right.\nonumber\\
+\left.\frac{\sqrt{3}v\hbar}{(\sqrt{3}v\hbar+ir(ae\mathbf{E}))}\right\}
\end{eqnarray}
and\\
\begin{eqnarray}
 j^{ac}=i\frac{4e^2 \gamma_0}{\sqrt{3}n\hbar^2} \sum_{r=1}r \sum_{s=1}{f_{rs}\varepsilon_{rs}}\sum_{m,l=-\infty}
{\frac{i^l mlj_m(\beta)j_{m-l}(\beta)I_{m-l}(\beta)(ae)\mathbf{E}}{(\sqrt{3}[\omega + iv]\hbar - m(ae)\mathbf{E})}}\nonumber\\
\times\sum_{r}\frac{\sqrt{3}v\hbar}{\sqrt{3}v\hbar+ir(ae\mathbf{E})}
\end{eqnarray}
where\\
\begin{eqnarray*}
f_{rs}=\frac{a}{2\pi\Delta p_{\vartheta}}\int_{0}^{2\pi/a}
\frac{\exp(-iarp/\sqrt{3})}{1+\exp\{\varepsilon_{s}(p)/k_{B}T\}}dp
\end{eqnarray*}
\begin{eqnarray*}
\varepsilon_{rs} = \frac{a}{2\pi\gamma_{0}}\int_{0}^{2\pi/a}
\varepsilon_{s}(\mathbf{p})\exp(iarp/\sqrt{3})dp
\end{eqnarray*}
$\beta =\kappa\gamma_{0}a/\Omega\sqrt{3}\hbar$, $\eta = Q/\Omega n_{0}$ and 
$\Omega=ea\mathbf{E}/\sqrt{3}\hbar$
\section*{Results and discussion}
We display the behaviour of the normalized current density ( $J = j/jos$) and 
$jos = (4\sqrt{3}e^{2}\gamma_0)/n\hbar$ ($zz$-CNT) or $4 e^{2}\gamma_{0}/\sqrt{3}n\hbar^{2}
(ac-CNT )$ as a function of the applied dc field $\mathbf{E}$ when frequency of $ac$ 
field $\omega$ is  much less than scattering frequency $v$($\omega\ll v$ or 
$\omega\tau\ll1$ i.e. quasi-static case) for the CNTs stimulated axially with the hot 
electrons, represented by the nonequilibrium parameter $\eta$ in figure $1$. 
\begin{figure}[h!]
\begin{centering}
\includegraphics[width = 12cm]{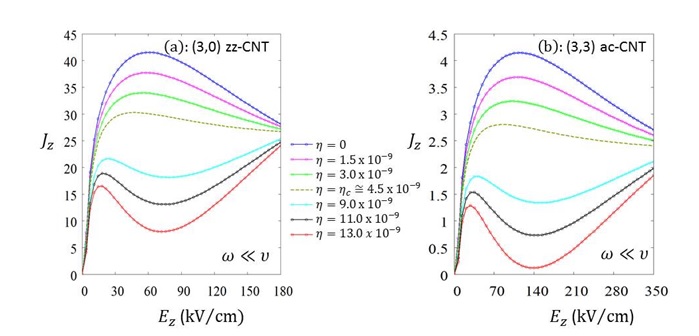} 
 \caption{A plot of  normalized current density  ($J_z$)  versus  applied dc  field 
($E_z$)  as the onequilibrium parameter $\eta$ increases from $0$ to 
$13.0 \times 10^{-9}$ when $\omega << v$ or 
$\omega\tau << 1$ (i.e.quasi-static case),  for (a) ($3,0$) zz-CNT  and 
(b) ($3,3$) ac-CNT, $T = 287.5K$ , $\omega = 10^{-4}  THz $, $v = 1 THz$ or 
$\tau = 1 ps$  and $\omega\tau=10^{-4}$} 
\end{centering}
\end{figure}
As we increase the nonequilibrium parameter $\eta$ from $0$ to $13.0\times10^{-9}$, 
we noticed that the normalized current density has the highest peak for $\eta$ (no 
hot electrons). As the hot electrons injection rate increases, the peak of the current 
density decreases and shifts to the left (i.e., low $dc$ fields). This is caused by the 
scattering effects due to electron-phonon interactions as well as the increase in the 
direct hot electrons injection rate~\cite{36}~\cite{37}. The normalized current density 
($J$) of the CNTs exhibits a linear monotonic dependence on the applied $dc$ field 
($\mathbf{E}$) at weak field (i.e.,the region of ohmic conductivity) when frequency of 
$ac$ field $\omega$ is  much less than scattering frequency $v$ ($\omega\ll v$ or 
$\omega\tau\ll1$ i.e. quasi-static case, where $v=\tau^{-1}$). As the applied $dc$ 
field ($\mathbf{E}$)  increases, the normalized current density ($J$)  increases and 
reaches a maximum, and drops off, experiencing a negative differential conductivity 
(NDC) for both the $zz$- CNT and the $ac$-CNT as shown in figures $1a$ and $1b$, 
respectively. The NDC is due to the increase in the collision rate of the energetic 
electrons with the lattice that induces large amplitude of oscillation in the lattice, 
which in-turn increases the electrons scattering rate that leads to the decrease in 
the current density at high dc field~\cite{37}. 
Similar effect was observed by Mensah, et. al.~\cite{40} in superlattice.
As the injection rate of the hot 
electrons becomes strong enough , the current density up-turned, exhibiting a 
positive differential conductivity (PDC) near $75kV/cm$ and $140kV/cm$ for the 
$zz$-CNT and the $ac$-CNT, respectively. In this region, the hot electrons become 
the dominant determining factor~\cite{36}. The physical mechanism behind the switch 
from NDC to PDC is due to the interplay between the hot electrons pumping frequency 
($Q/n_0$), which is a function of rate of hot electrons injection ($Q$), and the Bloch 
frequency ($\Omega$), which depends on the $dc$  field ($\mathbf{E}$)~\cite{37}. At 
stronger $dc$ field, the rate of scattering of the electrons by phonons is well 
pronounced resulting in the gradual decrease in the current density with increasing  
dc field (NDC region). However, as the rate of hot electrons injection increases, the 
corresponding rise in the current density due to hot electrons injection now far 
exceeds the reduction in the current density due to scattering of electrons by 
phonons. Thus, the net effect on the current density from the two opposing sources 
(with the hot electrons being dominant) gives rise to the PDC characteristics as 
shown in figure $1$ for $\eta \ge 9.0\times10^{-9}$. The desirable effect of a switch 
from NDC to PDC takes place when $\eta$ is larger than a critical value 
$\eta_c\approx 4.5\times 10^{-9}$. When axial injection of hot electrons into achiral 
CNTs is strong enough, the nonequilibrium parameter $\eta$ exceeds the critical value 
$\eta_c\approx 4.5\times 10^{-9}$ and the NDC characteristics change to the PDC 
characteristics. Thus, the most important tough problem for NDC region which is 
the space charge instabilities that inevitably lead to electric field domains 
formation resulting in non uniform electric field distribution which usually 
destroys THz Bloch gain can be suppressed due to the switch from the NDC 
behaviour to the PDC behaviour~\cite{38}. This is mainly due to the fact 
that PDC is considered as one of the conditions for electric stability of the 
system necessary for suppressing electric field domains ~\cite{38}. Hence a 
critical challenge for the successful observation of THz Bloch gain is the 
suppression of electric field domains by switching from NDC region to PDC region.
This is similar to that observed by Mensah, et. al.~\cite{40} in effect of ionization
of impurity centers in superlattice.
To put the above observations in perspective, we display in figure $2$, a 
$3$-dimensional behaviour of the normalized current density ($J$) as a function 
of the applied $dc$ field ($\mathbf{E}$) and nonequilibrium parameter ($\eta$) 
when frequency of $ac$ field $\omega$ is  much less than scattering frequency 
$v$ ($\omega\ll v$ or $\omega\tau\ll 1$ i.e. quasi-static case, where $v = \tau^{-1}$) 
for the CNTs.
\begin{figure}[h!]
\begin{centering}
\includegraphics[width = 12cm]{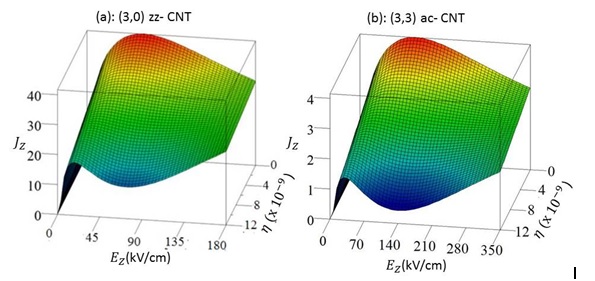} 
 \caption{A $3D$ plot of  normalized current density  ($J_z$)  versus  applied dc  field ($E_z$)  as the  
onequilibrium parameter $\eta$ increases for (a) ($3,0$) zz-CNT  and (b) ($3,3$) ac-CNT, 
when $\omega << v$ or $\omega\tau << 1$ (i.e. quasi-static case),   
$v = \tau^{-1}$, $T = 287.5K$ , $\omega = 10^{-4}  THz $, $v = 1 THz$ or $\tau=1 ps$  and $\omega\tau=10^{-4}$ } 
\end{centering}
\end{figure}
The $dc$ differential conductivity and the peak of the current density are at the 
highest when the nonequilibruim parameter $\eta$ is zero. For both $zz$-CNT and 
$ac$-CNT, as the nonequilibrium parameter $\eta$ gradually increases the $dc$ 
differential conductivity and the peak normalized current density decrease 
until the critical nonequilibrium parameter value $\eta_c\approx 4.5\times10^{-9}$ 
is reached, beyond which the NDC characteristics slowly changes to PDC characteristics 
as shown in figure 2. 
We further  display the behaviour of the normalized current density ($J$)  as a function of 
the applied $dc$ field ($\mathbf{E}$) of $ac-dc$ driven fields as $\omega\tau$ 
incrreasing from $0.01$ to $0.15$ when the nonequilibrium parameter $\eta = 0.9\times10^{-9}$ 
(presence of hot electrons) and $\eta=0$(absence of hot electrons) for $(3,0)$ 
$zz$- CNT and $(3,3)$ $ac$-CNT in figure $3$. 
\begin{figure}[h!]
\begin{centering}
\includegraphics[width = 12cm]{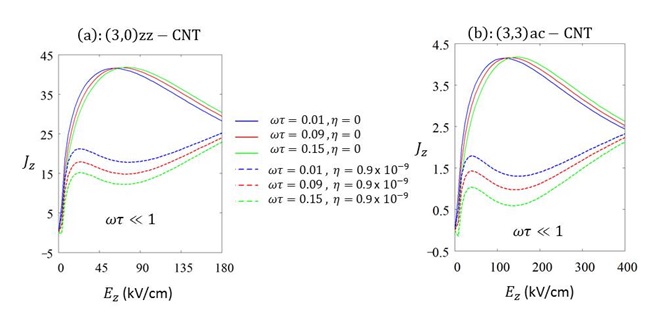} 
 \caption{A plot of  normalized current density  ( $J_z$)  versus  applied dc  field ($E_z$)  as 
$\omega\tau << 1$  increases from $0.01$ to  $0.17$ for (a) ($3,0$) zz-CNT  and (b) ($3,3$) ac-CNT  
when $\eta = 0$  and $\eta = 0.9 \times 10^{-9}$,   $v = 1 THz$ or $\tau = 1ps$} 
\end{centering}
\end{figure}
As we increase $\omega\tau$ from $0.01$ to  $0.15$ , we observed that the normalized 
current density has the highest peak at $\omega\tau=0.01$. Upon increasing the 
$\omega\tau$, the peak current density decreases until the least peak is attained 
when $\omega\tau=0.15$. Furthermore,we observed a switch from NDC to PDC  near  
$75kV/cm$ and $140kV/cm$  for $zz$-CNT and $ac$- CNT respectively so far as 
$\omega\tau\ll1$( i.e $0.01$ to $0.15$). Also the differential conductivity 
($\partial J/\partial\mathbf{E}$) in NDC region is fairly constant as $\omega\tau$ 
increases from $0.01$ to $0.15$. However in PDC region after the switch from NDC, 
differential conductivity($\partial J/\partial\mathbf{E}$)fairly increases as 
$\omega\tau$ increases from $0.01$ to $0.15$ as shown in figure $3a$ and $3b$ for 
zz-CNT and ac-CNT respectively. In the absence of hot electrons ($\eta=0$), we observed 
a shift of peak current density towards right ( i.e high dc-field) as $\omega\tau$ 
increases from $0.01$ to $0.15$ for each achiral CNT. Hence, the current
density dc field ($J-\mathbf{E}$)characteristics  for CNTs show a negative 
differential conductivity at stronger electric field without hot electrons and with 
strong enough axial injection of hot electrons (i.e. $\eta\ge0.9\times10^{-9}$), 
there is a switch from NDC to PDC leading to high electric field domain suppression 
necessary for generation of THz radiations provided $\omega\tau\ll1$(i.e quasi-static 
ac field).
\section*{Conclusion}
In summary, we have analyzed theoretically that strong enough injection of hot 
electrons in a CNT under conditions where, in addition to the dc field causing 
NDC, a similarly ac field is applied with a frequency $\omega$ much less than 
that of the scattering frequency $v$ (i.e. $\omega\ll v$ or $\omega\tau\ll1$, 
quasi-static case, $v=\tau^{-1}$), NDC switches to PDC. Hence, strong enough 
axial injection of hot electrons in CNT under the influence of quasi-static 
ac field  results in a switch from NDC to PDC leading to the suppression of  
the destructive electric domain instability, predicting a potential generation 
of terahertz radiations whose applications are  relevance in  current-day 
technology, industry, and research. Although similarly effect has been observed 
in the absence of quasi-static ac field [37], the differential conductivity 
($\partial J/\partial\mathbf{E}$) is higher  and also hot electrons injection 
rate beyond which there is a switch from NDC to PDC represented by critical 
noneqilibrium parameter $\eta_c$) is  lower in the presence of quasi-static 
$ac$- field than in the absence.

\end{document}